\documentclass[twocolumn,showpacs,preprintnumbers,amsmath,amssymb]{revtex4}

\usepackage{graphicx}
\usepackage{dcolumn}
\usepackage{bm}
\usepackage{color}

\begin{document}


\title{Jahn-Teller Inactivity and Magnetic Frustration in GeCo$_2$O$_4$ Probed by Ultrasound Velocity Measurements}

\author{Tadataka Watanabe$^1$}
\author{Shigeo Hara$^{2}$}
\author{Shin-Ichi Ikeda$^{2}$}
\affiliation{$^1$Department of Physics, College of Science and Technology (CST), Nihon University, Chiyoda-ku, Tokyo 101-8308, Japan \\
$^2$Nanoelectronics Research Institute, National Institute of Advanced Industrial Science and Technology (AIST), Tsukuba, Ibaraki 305-8568, Japan}
\date{\today}

\begin{abstract}
Ultrasound velocity measurements of cubic spinel GeCo$_2$O$_4$ in single crystal were performed for the investigation of shear and compression moduli. The shear moduli in the paramagnetic state reveal an absence of Jahn-Teller activity despite the presence of orbital degeneracy in the Co$^{2+}$ ions. Such a Jahn-Teller inactivity indicates that the intersite orbital-orbital interaction is much stronger than the Jahn-Teller coupling. The compression moduli in the paramagnetic state near the N$\acute{e}$el temperature $T_N$ reveal that the most relevant exchange path for the antiferromagnetic transition lies in the [111] direction. This exchange-path anisotropy is consistent with the antiferromagnetic structure with the wave vector $q \parallel$ [111], suggesting the presence of bond frustration due to competition among a direct ferromagnetic and several distant-neighbors antiferromagnetic interactions. In the JT-inactive condition, the bond frustration can be induced by geometrical orbital frustration of $t_{2g}$-$t_{2g}$ interaction between the Co$^{2+}$ ions which can be realized in the pyrochlore lattice of the high spin Co$^{2+}$ with $t_{2g}$-orbital degeneracy. In GeCo$_2$O$_4$, the tetragonal elongation below $T_N$ releases the orbital frustration by quenching the orbital degeneracy.
\end{abstract}

\pacs{71.70.Ej, 71.70.Gm, 75.10.-b, 75.50.Ee}

\maketitle

\section{Introduction}
Frustrated magnet systems have been a topical subject for more than a decade in condensed matter physics. Experimental and theoretical studies of geometrical spin frustration have revealed a variety of novel ground states such as spin ice and spin liquid states \cite{Moessner}. In addition to the spin degrees of freedom, the orbital degrees of freedom in $d$ electrons of the transition-metal ions can expand variations of the frustration effect and the exotic ground state. The cubic spinel structure $AB_2$O$_4$ provides a fertile field for the orbital physics in the highly magnetically frustrated network. The roles of the orbital sector analogous to the spin sector are under active discussions in the spinels. Very recently, for instance, the possibilities of orbital ordering \cite{Radaelli, Suzuki, Marita}, orbital liquid state \cite{Khomskii,Buttgen}, and orbital glassy state \cite{Fichtl} have been studied extensively.

The $B$-site sublattice of corner sharing tetrahedra in the spinels $AB_2$O$_4$,  so-called pyrochlore lattice, forms the most highly frustrated network. Thus the spinel structure with Jahn-Teller (JT) ions on the $B$ sites is the most suitable system to explore the frustration effects due to the interplay of the spin and the orbital degrees of freedom. The germanium-based spinel compound GeCo$_2$O$_4$ consists of magnetic Co$^{2+}$ (3$d^7$) ions on the octahedral $B$ sites with non-magnetic Ge$^{4+}$ ions on the tetrahedral $A$ sites. The $B$-site Co$^{2+}$ ions form the pyrochlore network as illustrated in Fig. 1 (a). In GeCo$_2$O$_4$, an antiferromagnetic (AF) transition occurs at the N$\acute{e}$el temperature $T_N$ = 23.0 K accompanied with a cubic to tetragonal structural elongation \cite{Hubsch1, Diaz1, Hoshi}. Magnetic susceptibility in the paramagnetic state, on the other hand, exhibits a Curie-Weiss behavior with the positive Weiss temperature $\Theta_W$ = 81.0 K indicating the dominant contribution of the ferromagnetic (FM) interactions \cite{Diaz1}. Recent experimental results of neutron powder diffraction suggest the presence of bond frustration due to competition among several FM and AF interactions in the AF state \cite{Diaz2}.

\begin{figure}[b]
\begin{center}
\includegraphics[height=70mm]{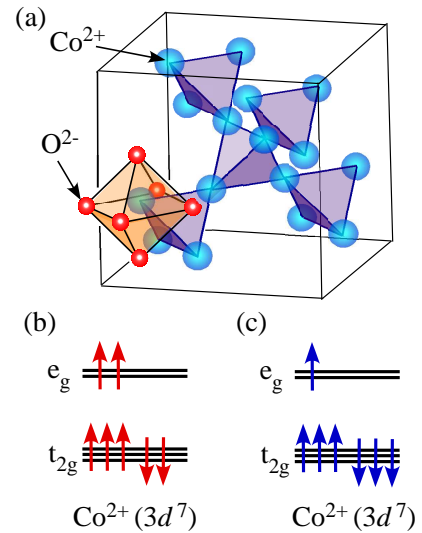}
\caption{\label{fig:Fig1} (Color online). (a) Pyrochlore lattice of the $B$-site Co$^{2+}$ ions in cubic spinel GeCo$_2$O$_4$. Octahedral O$^{2-}$ ligands surrounding a $B$ site Co$^{2+}$ ion are also illustrated. (b) High spin state, and (c) Low spin state of Co$^{2+}$ (3$d^7$) in the octahedral crystal field.}
\end{center}
\end{figure}

In GeCo$_2$O$_4$, the Co$^{2+}$ ion on the octahedral $B$ site should exhibit orbital degeneracy regardless the configuration of the 3$d^7$ electrons in the octahedral crystal field. The possible high spin and low spin states of the Co$^{2+}$ (3$d^7$) in the octahedral crystal field are depicted in Fig. 1 (b) and (c), respectively. There exists threefold degeneracy of $t_{2g}$ orbitals in the high spin state (Fig. 1 (b)), while twofold degeneracy of $e_g$ orbitals in the low spin state (Fig. 1 (c)). We can expect, therefore, that the orbital degrees of freedom play significant roles in the magnetism of GeCo$_2$O$_4$. Although the small frustration factor $f=|\Theta_{W}/T_N| \simeq 3.5$ rules out the possibility of the simple geometrical spin frustration in this compound, it is worthwhile to explore another kind of frustration due to the interplay of the spin and the orbital degrees of freedom. Orbital physics in GeCo$_2$O$_4$, however, has not yet been studied both experimentally and theoretically.

In this paper, we present ultrasound velocity measurements in GeCo$_2$O$_4$ single crystals in shear and compression moduli which provide angle-resolved information of the elastic properties. Ultrasound velocity is a powerful tool for the study of orbital physics because electric quadrupole moments of JT ions strongly couple to the lattice deformations \cite{Ishikawa, Kino, Kataoka, Hazama}. In addition, since ultrasound wave modulates the distance between magnetic ions, ultrasound velocity is also a directional probe of the anisotropy of exchange interactions acting on the magnetic phase transition \cite{Kawasaki}. The present study discusses orbital state, spin state, and possible magnetic frustration in GeCo$_2$O$_4$ from the elastic properties.

\section{Experimental}
Large single crystals of GeCo$_2$O$_4$ were grown by the floating zone method. Details of the crystal growth are described elsewhere  \cite{Hara}. The N$\acute{e}$el temperature $T_N$ and the Weiss temperature $\Theta_W$ of the single crystals applied for the present study were determined by the magnetic susceptibility measured by a SQUID magnetometer. Figure 2 depicts the temperature dependence of the inverse magnetic susceptibility with $\mu_0 H$= 1 T. The antiferromagnetic transition occurs at $T_N$ = 20.6 K, and a linear fit to the experimental data above 160 K yields the Weiss temperature $\Theta_W$ = 41.0 K, leading to the small frustration factor $f=|\Theta_{W}/T_N| \simeq 1.99$. The discrepancy of $\Theta_W$ between the present study ($\Theta_W$ = 41.0 K) and Ref. 9 ($\Theta_W$ = 81.0 K) would be attributed to the difference of the temperature range of the linear fitting: 160 K $<T<$ 300 K in the present study, while 300 K $<T<$ 800 K in Ref. 9. In fact, $\Theta_W$ in Ref. 8 determined by the fitting in the temperature range 160 K $<T<$ 300 K exhibits $\Theta_W$ = 47.6 K which is comparable to the present study. Extension of the fitting range to higher temperatures in the present single crystals would yield more reliable and higher $\Theta_W$.

\begin{figure}[t]
\begin{center}
\includegraphics[height=55mm]{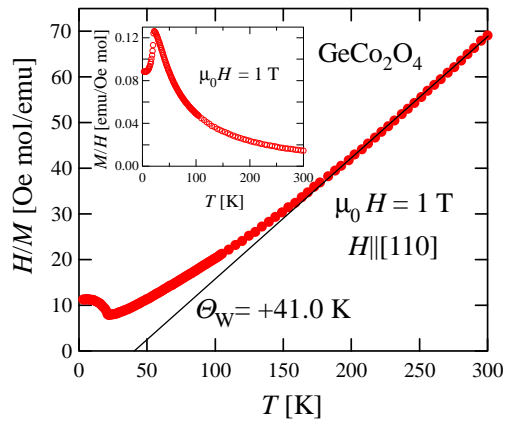}
\caption{\label{fig:Fig2} (Color online). Inverse magnetic susceptibility as a function of temperature with $\mu_0 H$ = 1 T. Inset shows the temperature dependence of the magnetic susceptibility with $\mu_0 H$ = 1 T.}
\end{center}
\end{figure}

For the ultrasound measurements, the crystal was cut into cuboid shape with the dimension of 3.6$\times$3.5$\times$2.6 mm$^3$. Mirror surfaces of the sample were prepared by careful polishing using the 1$\mu$m diamond slurry because the ultrasound measurements are quite sensitive to the roughness of the sample surface. The ultrasound velocity measurements were performed by a home-built apparatus with phase comparison technique. This apparatus can measure the relative change of the ultrasound velocities with high resolution of about 1 part in 10$^6$. For the generation and the detection of ultrasounds, we used LiNbO$_3$ transducers with the fundamental frequency of 30 MHz. The transducers were glued on the parallel mirror surfaces of the sample by silicone adhesive. Ultrasound velocities were measured in the temperature range from 2 K to 100 K with the magnetic field from 0 to 7 T.

\begin{table}[b]
\caption{\label{tab:table1} Ultrasound modes investigated in this study and the corresponding sound propagation direction \boldmath$k$, polarization \boldmath$u$, and elastic modulus in cubic symmetry.}
\begin{ruledtabular}
\begin{tabular}{cccc}
Ultrasound mode& \boldmath$k$ & \boldmath$u$ & Elastic modulus\\
\hline
Transverse wave 1 & [001] & [110] & $C_{44}$ \\
Transverse wave 2 & [110] & [1$\bar{1}$0] & $\frac{C_{11}-C_{12}}{2}$ \\
\hline
Longitudinal wave 1 & [100] & [100]& $C_{11}$\\
Longitudinal wave 2 & [110] & [110]& $\frac{C_{11}+C_{12}+2C_{44}}{2}$\\
Longitudinal wave 3 & [111] & [111]& $\frac{C_{11}+2C_{12}+C_{44}}{2}$\\
\end{tabular}
\end{ruledtabular}
\end{table}

The ultrasound velocities were measured in five different ultrasound modes in a GeCo$_2$O$_4$ single crystal. TABLE I summarizes the ultrasound modes and the corresponding propagation direction \boldmath$k$, polarization direction \boldmath$u$, and elastic modulus investigated in the present study. In the crystal structure with the cubic symmetry, there are three elastic moduli which are symmetrically independent: $C_{11}$, $C_{12}$, and $C_{44}$. We measured transverse ultrasound velocities in tetragonal shear modulus $\frac{(C_{11}-C_{12})}{2}$ with $\Gamma_3 (E_g)$ symmetry, and trigonal shear modulus $C_{44}$ with $\Gamma_5 (T_{2g})$ symmetry. Longitudinal ultrasound velocities were measured in three compression moduli with the different directions of \boldmath$k$ (\boldmath$u$) along $[100]$, $[110]$, and $[111]$ directions. Motivations of the measurements of the transverse and the longitudinal ultrasound velocities are described with their experimental results in Sec. III A and B, respectively.

\section{Results And Discussion}
\subsection{Shear moduli}
First we investigate the JT activity in GeCo$_2$O$_4$ focusing on the orbital degrees of freedom. As depicted in Fig. 1 (b) and (c), the Co$^{2+} (3d^7)$ on the octahedral $B$ site contains the orbital degrees of freedom regardless the spin state: threefold-degenerate $t_{2g}$ orbitals in the high spin state ($S=3/2$), while twofold-degenerate $e_g$ orbitals in the low spin state ($S=1/2$). If the cubic-to-tetragonal structural transition at $T_N$ in GeCo$_2$O$_4$ originates from the cooperative JT effect, the tetragonal shear modulus $\frac{(C_{11}-C_{12})}{2}$ with $\Gamma_3 (E_g)$ symmetry strongly and selectively couples to the Co$^{2+}$ ions via the quadrupole-strain interaction written as \cite{Hazama, Kataoka, KK4},

\begin{equation}
H_{QS}= -\sum_{i}g_{\Gamma_3}O^2_2(i) \epsilon_{v}.
\label{eq:QS}
\end{equation}
\\
Here $g_{\Gamma_3}$ is the coupling constant, $O^2_2(i)=l_x^2-l_y^2$ is the quadrupole operator of the $i$th site expressed by the components of the angular momentum $l_x$ and $l_y$, and $\epsilon_v=\epsilon_{xx}-\epsilon_{yy}$ is the elastic strain. Thus $\frac{(C_{11}-C_{12})}{2}$ corresponds to the elastic mode for the verification of JT activity in GeCo$_2$O$_4$. Taking into account the intersite quadrupole-quadrupole interaction, the quadrupole-strain interaction of Eq.~(\ref{eq:QS}) leads to the temperature dependence of the JT-active elastic modulus $C_{\Gamma_3}=\frac{C_{11}-C_{12}}{2}$ in the cubic phase $T \textgreater T_N$ written as \cite{Kino, Kataoka, Hazama},

\begin{equation}
C_{\Gamma_3} = C^0_{\Gamma_3} \frac{T-(\Theta+E_{JT})}{T-\Theta},
\label{eq:JT}
\end{equation}
\\
with $C^0_{\Gamma_3}$ the elastic constant without JT effect, $\theta$ the intersite orbital-orbital (quadrupole-quadrupole) interaction, $E_{JT}$ the JT coupling energy. In the scenario of the cooperative JT effect, it is expected from Eq.~(\ref{eq:JT}) that, in GeCo$_2$O$_4$, $\frac{C_{11}-C_{12}}{2}$ exhibits a precursor to the structural phase transition: a huge softening with decreasing temperature in wide temperature range of $T \textgreater T_N$ \cite{Ishikawa, Kino, Kataoka, Hazama}.

\begin{figure}[t]
\begin{center}
\includegraphics[height=55mm]{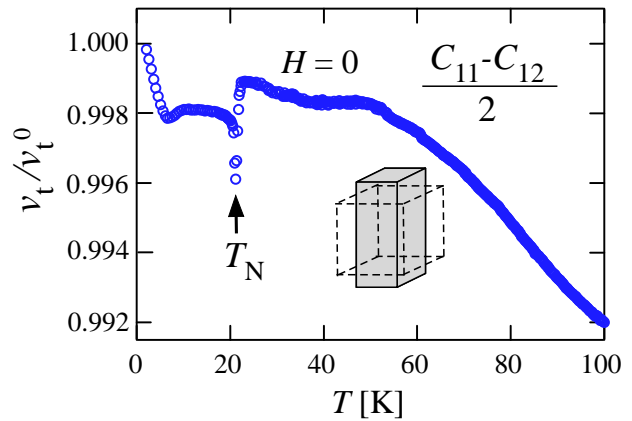}
\caption{\label{fig:Fig2} (Color online). The transverse ultrasound velocity $v_t$ in tetragonal shear modulus $\frac{(C_{11}-C_{12})}{2}$ as a function of temperature with $H=0$. Inset picture shows schematic of the tetragonal lattice deformation in $\frac{(C_{11}-C_{12})}{2}$ ($E_g$ symmetry).}
\end{center}
\end{figure}

Figure 3 depicts the temperature dependence of the transverse ultrasound velocity $v_t$ in $\frac{(C_{11}-C_{12})}{2}$ in zero magnetic field. In the paramagnetic state, $\frac{(C_{11}-C_{12})}{2}$ exhibits ordinary hardening with decreasing temperature, instead of the softening expected from the JT effect. In GeCo$_2$O$_4$, the orbital-orbital interaction and the JT coupling should be present due to the orbital degeneracy of Co$^{2+}$ as depicted in Fig. 1 (b) and (c): $\Theta\neq0$ and $E_{JT}\neq0$ in Eq.~(\ref{eq:JT}). Thus the observed elastic property of $\frac{(C_{11}-C_{12})}{2}$ in the paramagnetic state reveals that the JT coupling is negligibly weak compared to the orbital-orbital interaction, $\Theta \gg E_{JT}\neq0$ in Eq.~(\ref{eq:JT}), where the elasticity behaves like $C_{\Gamma}\simeq C^0_{\Gamma}$. In the vicinity of $T_N$, a dip-like anomaly is seen due to the phase transition. In the antiferromagnetic (AF) state, the elasticity steeply hardens below $\sim$ 6 K down to the lowest temperature. As will be described below in conjunction with Fig. 5, this might be attributed to the relaxation of the stress on the magnetic domain walls.

\begin{figure}[b]
\begin{center}
\includegraphics[height=67mm]{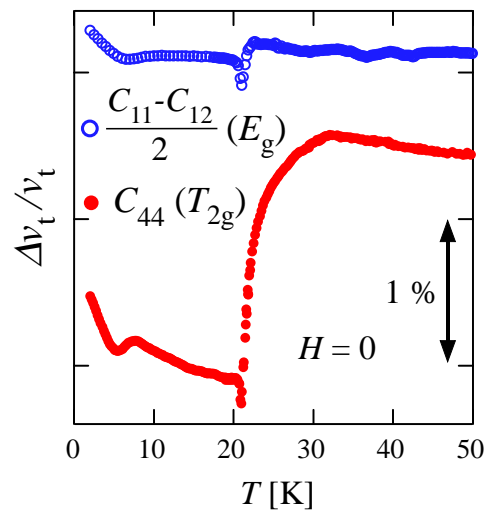}
\caption{\label{fig:Fig3} (Color online). Relative change of the transverse ultrasound velocity $v_t$ with $H=0$ as a function of temperature in $\frac{(C_{11}-C_{12})}{2}$ ($E_g$ symmetry), and $C_{44}$ ($T_{2g}$ symmetry). For clarity, the zero levels of the different curves are vertically shifted.}
\end{center}
\end{figure}

The relative change of the trigonal shear modulus $C_{44}$ as a function of temperature in zero magnetic field is presented in Fig. 4 compared on the same scale with the result of $\frac{(C_{11}-C_{12})}{2}$ shown in Fig. 3. Similar to $\frac{(C_{11}-C_{12})}{2}$, $C_{44}$ exhibits a dip-like anomaly of $\Delta v_t/v_t \sim0.2 \%$ in the vicinity of $T_N$. Figure 5 (a) and (b) show the relative shift of $\frac{(C_{11}-C_{12})}{2}$ and $C_{44}$, respectively, with magnetic fields \boldmath$H \|$[110] as a function of temperature. The dip-like anomalies in both $\frac{(C_{11}-C_{12})}{2}$ and $C_{44}$ shift to lower temperatures with increasing $H$ following the reduction of $T_N$. This correspondence indicates that the dip-like elastic anomalies are relevant to the AF transition. Such an elastic anomaly is attributed to the magneto-elastic coupling acting on the exchange interactions \cite{Kawasaki}. In this mechanism, the exchange striction arises from a modulation of the exchange interactions by ultrasound as follows,

\begin{equation}
H_{exs} = \sum_{ij}[J(\delta + u_i - u_j) - J(\delta)]S_i \cdot S_j.
\label{eq:ES1}
\end{equation}
Here $\delta = R_i-R_j$ is the distance between two magnetic ions, and $u_i$ is the displacement vector for the ion $R_i$.

In Fig. 4 and Fig. 5 (b), a salient anomaly in $C_{44}$ is the $\sim1.5 \%$ steep softening from $\sim$30 K down to $T_N$. As shown in Fig. 5 (b), this softening shifts to lower temperatures with increasing $H$ following the reduction of $T_N$, indicative the relevance of this anomaly to the AF transition. Thus this elastic anomaly is also attributed to the exchange-striction effect of Eq.~(\ref{eq:ES1}). Such an elastic instability observed only in $C_{44}$ suggests, with its absence in $\frac{C_{11}-C_{12}}{2}$, the presence of highly anisotropic magnetic fluctuations in GeCo$_2$O$_4$. This $C_{44}$-selective anomaly is in accordance with the results of neutron scattering experiments suggesting the presence of two-dimensional magnetic fluctuations from $\sim$25 K down to $T_N$ \cite{Hubsch}.

\begin{figure}[b]
\begin{center}
\includegraphics[height=60mm]{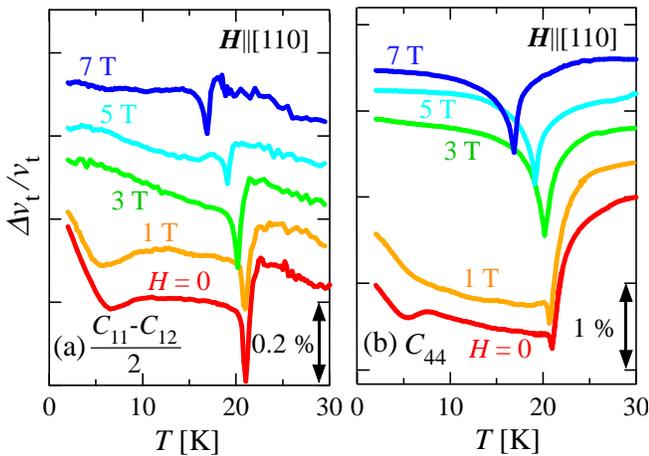}
\caption{\label{fig:Fig4} (Color online). Relative shift of the transverse ultrasound velocity $v_t$ with magnetic field $H \|$[110] as a function of temperature in (a) $\frac{C_{11}-C_{12}}{2}$, and (b) $C_{44}$. For clarity, the zero levels of the different curves are vertically shifted.}
\end{center}
\end{figure}

Here, we would also like to comment on the elasticities in the AF state shown in Fig. 5, although the tetragonal distortion of the crystal in the AF state transforms the elastic symmetry. In the AF state without $H$ or with low $H$ less than 1 T, $C_{44}$ shown in Fig. 5 (b) exhibits weak temperature dependence with loose hardening from just below $T_N$ down to $\sim$8 K where the magnitude of the detected ultrasound is extremely attenuated. $C_{44}$ with $H$ higher than 1 T, in contrast, promptly hardens below $T_N$ with moderate attenuation of the ultrasound. These $H$-dependent elasticities in the AF state would be attributed to the stress effect on the magnetic domain walls \cite{Luthi}. Magnetization of the sample into a single domain state with $H$ can recover the elasticity and the magnitude of the ultrasound as observed in the present experiments. The enhancement of the ultrasound attenuation in the AF state with low $H$ less than 1 T is also observed in $\frac{(C_{11}-C_{12})}{2}$, but is not so large compared to $C_{44}$. In Fig. 5 (a) and (b), $\frac{(C_{11}-C_{12})}{2}$ and $C_{44}$ in $H$ = 0 (for $\frac{(C_{11}-C_{12})}{2}$, also in $\mu_0$\boldmath$H$ = 1 T) harden steeply below $\sim$ 6 K down to the lowest temperature with the recoveries of the magnitude of the ultrasounds. This hardening might be attributed to the formation of the single domain state even in zero magnetic field. The origin of this zero-field hardening is not clear at present, and further studies should be performed for the elucidation.

\subsection{Compression moduli}
The experimental results in the shear moduli described above revealed that the dominant mechanism of the observed elastic anomalies in GeCo$_2$O$_4$ is the exchange-striction effect expressed by Eq.~(\ref{eq:ES1}). Next, we study more in detail the anisotropy of the exchange interactions acting on the AF transition by investigating the compression moduli with the longitudinal ultrasounds. In the exchange-striction mechanism, the longitudinal sound wave exhibits more directional feature than the transverse sound wave. When a sound wave with polarization $u$ and propagation $k$ is given by $u$ = $u_0$exp[i($k \cdot r-\omega t$)], where $u_0$ and $\omega$ are amplitude and frequency respectively, the exchange striction of Eq.~(\ref{eq:ES1}) is rewritten as \cite{Stern},

\begin{equation}
H_{exs} = \sum_{i}(\frac{dJ}{d\delta} \cdot u)(k \cdot \delta)(S_i \cdot S_{i+\delta})e^{i(k \cdot R_i-\omega t)}.
\label{eq:ES2}
\end{equation}
The longitudinal sound wave, therefore, strongly couples to the exchange interactions acting in the direction parallel to the sound propagation $k$ and polarization $u$, $k\parallel\delta$ and $u\parallel\delta$.

\begin{figure}[t]
\begin{center}
\includegraphics[height=75mm]{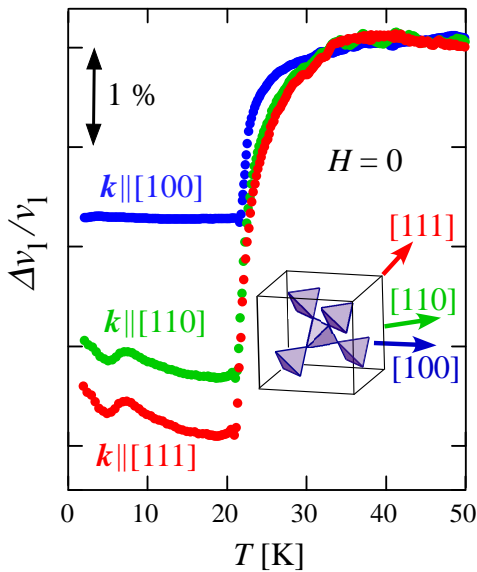}
\caption{\label{fig:Fig5} (Color online). Relative change of the longitudinal ultrasound velocity $v_l$ as a function of temperature with $H=0$ in $k||[100]$, $k||[110]$, and $k||[111]$. Inset picture shows $k$ directions of each mode in the $B$-site pyrochlore lattice.}
\end{center}
\end{figure}

Figure 6 depicts the temperature dependence of the longitudinal ultrasound velocities with the propagation $k\parallel[100]$, $k\parallel[110]$, and $k\parallel[111]$ in zero magnetic field. All the compression moduli exhibit steep softening from $\sim$35 K down to $T_N$ due to the exchange-striction effect. Among these elastic moduli, the magnitude of the anomalous variation in $k\parallel[111]$ is the largest indicative the strongest contribution of the exchange interactions acting in the $[111]$ direction to the AF transition. This anisotropy of the exchange-striction effect in the paramagnetic state near $T_N$ is consistent with the AF structure with the wave vector $q\parallel[111]$ determined by the neutron scattering experiments \cite{Diaz2}. As pointed out in Ref. 11, the formation of the AF structure with $q\parallel[111]$ despite the positive value of the Weiss temperature $\Theta_W \textgreater 0$ suggests the magnetic frustration due to competition among several FM and AF interactions. The present results of the compression moduli in the paramagnetic state near $T_N$ strongly support the possibility of such a bond frustration in GeCo$_2$O$_4$. We discuss the origin of the bond frustration in Sec. III C. In the AF state, the compression moduli exhibit weak or almost no temperature dependence with the enhanced attenuation of the ultrasound signals. As already pointed out in the shear moduli in Sec. III A, these behaviors would be attributed to the stress effect on the magnetic domain walls.

\subsection{Spin state of Co$^{2+}$ and magnetic frustration in GeCo$_2$O$_4$}
Here, we would like to discuss the spin state of Co$^{2+}$ and magnetic frustration in GeCo$_2$O$_4$ from the elastic properties revealed by the present study. As depicted in Fig. 1 (b) and (c), the Co$^{2+}$ on the octahedral $B$ site contains the orbital degrees of freedom regardless the spin state: threefold-degenerate $t_{2g}$ orbitals in the high spin state ($S=3/2$), while twofold-degenerate $e_g$ orbitals in the low spin state ($S=1/2$). In the present experiments, however, the absence of softening in the tetragonal shear modulus $\frac{(C_{11}-C_{12})}{2}$ in the paramagnetic state (Fig. 3) excludes the possible JT-active mechanism in the cubic-to-tetragonal structural transition at $T_N$. Instead, as shown in Fig. 4, $\frac{(C_{11}-C_{12})}{2}$ and $C_{44}$ exhibit the dip-like anomalies in the vicinity of $T_N$ which are relevant to the AF transition. We conclude, therefore, that the cubic-to-tetragonal structural distortion is driven by the AF ordering.

The dip-like anomalies not only in $\frac{(C_{11}-C_{12})}{2}$ with $E_g$ symmetry but also in $C_{44}$ with $T_{2g}$ symmetry indicate that both $e_g$ and $t_{2g}$ orbitals of the Co$^{2+}$ participate in the exchange interactions. Comparing between the high spin state of Fig. 1 (b) and the low spin state of Fig. 1 (c), both $t_{2g}$ and $e_g$ orbitals can participate in the exchange interactions in the high spin state, whereas only the $e_g$ orbitals can participate in the low spin state. The dip-like anomalies in the shear moduli, therefore, lead to the conclusion that the Co$^{2+}$ in GeCo$_2$O$_4$ adopts high spin state in agreement with the magnetic susceptibility measurements of Ref. 9. Another salient anomaly of the steep softening observed only in $C_{44}$ below $\sim$30 K down to $T_N$ (Fig. 4) robustly supports this conclusion. Such an anomaly in $C_{44}$ suggests the presence of magnetic fluctuation sensitive to the trigonal lattice deformation with $T_{2g}$ symmetry. The steep softening only in $C_{44}$ indicates that the exchange interactions via $t_{2g}$ orbitals dominate the magnetic fluctuation. Such a magnetic fluctuation can be realized in the network of the high spin Co$^{2+}$ with $t_{2g}$-orbital degeneracy.

\begin{figure}[t]
\begin{center}
\includegraphics[height=82mm]{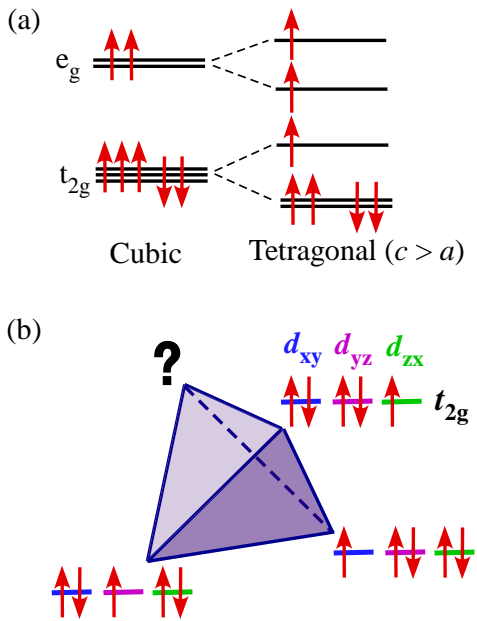}
\caption{\label{fig:Fig6} (Color online). (a) Crystal-field splitting by cubic-to-tetragonal structural transition with $c\textgreater a$ in the Co$^{2+}$ (3$d^7$). (b) Geometrical orbital frustration of the direct FM (Kugel-Khomskii-type) interaction between the $t_{2g}$ orbitals in the Co$^{2+}$ tetrahedron.}
\end{center}
\end{figure}

On the basis of the high spin Co$^{2+}$ concluded above, we next discuss the magnetic frustration in GeCo$_2$O$_4$. The JT inactivity in $\frac{(C_{11}-C_{12})}{2}$ despite the presence of the orbital degrees of freedom indicates that, as understood from Eq.~(\ref{eq:JT}), the orbital-orbital interaction $\Theta$ is much stronger than the JT coupling $E_{JT}$, $\Theta \gg E_{JT}$. Furthermore, the steep softening of $C_{44}$ in the paramagnetic state near $T_N$ indicates that the $t_{2g}$-$t_{2g}$ exchange interactions play a significant role in the magnetic fluctuation near $T_N$. According to the Goodenough-Anderson-Kanamori (GAK) rules, the presence of a hole in the $t_{2g}$ orbitals in Co$^{2+}$ gives rise to the direct FM interaction between the nearest-neighbor Co$^{2+}$ ions \cite{Goodenough}. This FM interaction is strong and consistent with the positive Weiss temperature $\Theta_W \textgreater 0$ in GeCo$_2$O$_4$. The anisotropy of the elastic anomaly in the compression moduli shown in Fig. 6, however, strongly suggests the presence of the bond frustration. The GAK rules tell that the bond frustration in GeCo$_2$O$_4$ arises from the competition among direct FM interaction, third-neighbor AF interaction, and sixth-neighbor AF interaction, as pointed out in Ref. 11.

The question is why the strong direct-FM interaction competes with the weak distant-neighbors-AF interactions. The most probable answer from the present study is the presence of geometrical orbital frustration in the paramagnetic state in GeCo$_2$O$_4$. In the $t_{2g}$-orbital degenerate system, Kugel-Khomskii-type (KK) ferromagnetic interaction \cite{KK1, Inagaki, KK2, KK3, KK4} can be more emphasized since partially filled $t_{2g}$ orbitals have greater degeneracy and weaker JT coupling compared to the $e_g$ orbitals. The JT inactivity with $\Theta \gg E_{JT}$ in Eq.~(\ref{eq:JT}) strongly suggests the dominant contribution of the orbital-orbital interaction. It is possible, therefore, that a FM transition with orbital ordering occurs in GeCo$_2$O$_4$ by the KK interaction. However, GeCo$_2$O$_4$ exhibits the AF transition, and, as depicted in Fig. 7 (a), the tetragonal elongation below $T_N$ quenches the $t_{2g}$-orbital degeneracy. Thus the possibility of the orbital ordering is ruled out in GeCo$_2$O$_4$. Instead, the tetragonal elongation shown in Fig. 7 (a) can release the geometrical orbital frustration in this compound. Figure 7 (b) illustrates the schematic picture of the $t_{2g}$-orbital frustration in the Co$^{2+}$ tetrahedron. When a single hole in $t_{2g}$ orbitals is aligned alternately in three Co$^{2+}$ sites of the tetrahedron by the KK interaction, another fourth Co$^{2+}$ site experiences the geometrical frustration for the hole (orbital) arrangement. This situation disturbs the occurrence of the orbital ordering. On the other hand, the quenching of the orbital degeneracy by the tetragonal elongation releases the orbital frustration. The magnetic fluctuation near $T_N$ where $t_{2g}$ orbitals dominantly contribute is probably a precursor to the release of the orbital frustration. It is noted that this AF transition with the tetragonal elongation is a new kind of phase transition driven by the orbital degeneracy which is different from the cooperative JT distortion and the orbital ordering.

\section{Summary}
To summarize, we performed ultrasound velocity measurements in single crystals of GeCo$_2$O$_4$ in shear and compression modes. The shear moduli reveal JT inactivity despite the presence of $t_{2g}$-orbital degeneracy in the high-spin Co$^{2+}$ ions. The compression moduli show anisotropic elastic anomalies near $T_N$ suggesting the presence of bond frustration due to the competition among direct FM and distant-neighbors AF interactions. The most probable origin of the bond frustration with the JT inactivity in GeCo$_2$O$_4$ is the presence of geometrical orbital frustration which can be realized in the pyrochlore lattice formed by JT ions with threefold-degenerate $t_{2g}$ orbitals. The cubic-to-tetragonal elongation by AF ordering releases the orbital frustration by quenching the orbital degeneracy. Further experimental and theoretical efforts are necessary for the verification of the orbital frustration in GeCo$_2$O$_4$.

\section{Acknowledgments}
We thank K. Tomiyasu for helpful comments and fruitful discussions.

\end{document}